# INSTABILITY OF THE FERROFLUID LAYER ON A MAGNETIZABLE SUBSTRATE IN A PERPENDICULAR MAGNETIC FIELD

A.R. Zakinyan, L.S. Mkrtchyan

*Institute of Natural Sciences, North-Caucasian Federal University, 355009 Stavropol, 1 Pushkin St., Russian Federation*

**Abstract.** The paper presents an experimental study of the instability of a magnetic fluid layer of finite thickness covering a magnetizable metal plate exposed to a perpendicular magnetic field. The critical field strength and the instability wave number have been measured.

## 1. Introduction

A ferrofluid or magnetic fluid is an ultra disperse colloidal solution of ferromagnetic or ferrimagnetic nanoparticles in a nonmagnetic fluid medium. Its capacity to effectively interact with magnetic field results in a number of physical effects; one of them is the instability of its surface in magnetic field. A classical example of such effect is the instability of the flat free surface of magnetic fluid in a uniform perpendicular magnetic field (Rosensweig instability) [1, 2]. When the external magnetic field strength reaches a certain critical value, the surface of the magnetic fluid stops being smooth and a structure of conical peaks forms on it in an intermittent manner. The surface structure development becomes specific when the fluid layer is of finite thickness, especially when this thickness is small compared with the wave length of the occurring disturbances.

Earlier, in [3], we presented a study of instability of the thin layer of magnetic fluid on nonmagnetic substrate under effect of perpendicular magnetic field. It has been reported that the most expressed dependence of layer instability development character on layer thickness was observed in the range of thicknesses of the order of several tens of micrometers. The magnetic fluid thick layer instability development results, as a rule, in that a system of conical peaks occurs on its surface; these peaks form a periodical hexagonal structure [1, 2]. As it has been shown in [3], the specificity of the thin layer instability development is the possibility of occurrence of a secondary hexagonal structure consisting of the smaller size peaks formed together with the primary structure peaks.

The regularities of occurrence of the instability of the magnetic fluid layer covering the substrate with strong magnetic properties may have a number of additional specificities. The available literature does not, practically, deal with these problems. There are several theoretical studies, e.g., [4–6], while some aspects of this issue have been studied experimentally only in [7]. Meanwhile, this situation may take place in a number of technical applications of magnetic fluids and its study is of both the general scientific and engineering interest. To fill this gap, in this work we experimentally study the instability of the flat horizontal, 5 µm to 50 µm thick, layers of a magnetic fluid covering a magnetizable plate, in uniform perpendicular magnetic field. We have also compared the results of this study with the ones obtained earlier when studying the instability of a fluid layer on nonmagnetic substrate.

## 2. Experimental and Results

The experiments studied the horizontal layer of ferrofluid applied on a flat plate made of transformer steel. The plate was 1 mm thick. The magnetic fluid used in experiments was a dispersion of magnetite nanoparticles, mean diameter 10 nm, in kerosene; the oleic acid was used as stabilizer. The magnetic fluid density was $\rho = 1400$ kg/m³, its saturation magnetization was $M_s = 55.4$ kA/m. The surface tension coefficient at the fluid-air interface was $\sigma = 0.028$ N/m. Fig. 1 shows the measured magnetization curves of the magnetic fluid and substrate material.

The experimental technique was as follows. A magnetic fluid layer applied on a metal plate ($S = 5$ cm²) was put between electromagnet poles, in a uniform magnetic field area. The fluid layer thickness $h$ was determined after the known volume $V$ of the fluid, uniformly distributed by the plate surface ($h = V/S$). Then, the magnetic field strength was smoothly increased; the condition of the magnetic field layer was observed visually.

When the field strength reached a certain critical value, conical peaks formed on the surface of the layer; their axes coincided with the magnetic field direction. The development of instability practically immediately provoked the layer disintegration into individual conical drops of magnetic fluid on the substrate; these drops tended to form a hexagonal system. The external magnetic field strength $H_c$ at which the fluid became instable was measured. The forming system of peaks was practically uniform on the whole surface of the layer, hence, in these experiments, the edge effects may be neglected (due to the small instability wave length compared with the layer plane size). The magnetic fluid layer structure formed due to the effect of the field was then dried and studied using an optical microscope. During drying (~ 15 minutes) the spatial periodicity of the formed structure remained unchanged. The mean distance between adjacent peaks was measured; the so found disturbance wave length $\lambda_c$ served to calculate the critical wave number $k_c = 2\pi/\lambda_c$.

Fig. 2 shows the experimental dependence of the external magnetic field critical strength $H_c$ at which the instability occurs upon the ferrofluid layer thickness. The figure shows that this dependence monotonically decreases. The figure also shows the earlier results corresponding to the magnetic field critical strength at which the ferrofluid layer on nonmagnetic substrate becomes instable. The instability of the layer applied on a magnetizable substrate can be seen to occur at the lower field strengths. In other words, the magnetic substrate stimulates the development of surface instability under effect of magnetic field.

Fig. 3 shows the experimental dependence of the critical wave number $k_c$ characterizing the occurring instability upon the fluid layer thickness. The results for a nonmagnetic substrate are shown, as well. The figure shows that in the relatively thick layers the dependences for magnetic and nonmagnetic substrates are practically same. A visible difference is observed only in the less than 15 µm thick layers. The conclusion can be made that as the magnetic fluid layer thickness grows the effect of the substrate properties on the instability development process fades away quickly enough, and the thick layers instability is practically independent of the layer substrate.

If the magnetic field strength continues growing above $H_c$, the peaks amplitude may grow, however, the formed structure and distances between peaks remain practically unchanged. However, the instability development character may essentially change if the magnetic field growth rate is greater than the instable disturbances increment. In this case, the instability may develop on the background of different field strengths resulting in changed instability parameters; in particular, the most unstable disturbances wave number $k$ depends in such event upon the applied field. Due to the above, the studies continued with the instantaneous switching of the magnetic field, a fortiori above the critical value corresponding to the start of the magnetic fluid layer instability development. In our experiments the field creation time was short enough ($< 0.5$ s), hence, the field may be considered stabilized before the disturbances develop.

It was found that at such abrupt application of magnetic field the occurring layer structure depends heavily upon the field strength. Thus, a small supercriticality of the field results in the classical structure of practically identical peaks (Fig. 4a). The field strength growth results in additional smaller peaks (Fig. 4b); see in [2] the description of the ferrofluid on nonmagnetic substrate layer instability study. However, in this case, the secondary peaks occur at lower field strengths than with a nonmagnetic substrate. The further field growth resulted in even smaller peaks, appearing together with the primary and secondary ones (Fig. 4c), not observed in the earlier nonmagnetic substrate experiments. The conclusion is possible that the tertiary and even quaternary peaks (Fig. 4c) are characteristic for the development of the instability of the thin layer of magnetic fluid on magnetizable substrate. It should be noted that the system of peaks forming on a magnetic substrate is less ordered than with a nonmagnetic one; in this case the hexagonal structure is practically unseen.

The wave number $k$ of the instability developing at the field supercritical strengths was measured. For this, the mean distance $\lambda$ between adjacent primary peaks was measured. The distances between the smaller secondary and tertiary peaks were not studied. Fig. 5 shows the experimental dependence of the instability supercritical wave number upon the external magnetic field strength compared with the relevant dependence for the layer on nonmagnetic substrate. It is seen that at weak fields the both dependences practically merge. But as the field grows, the dependence for the layer on magnetic substrate goes through an extremum while that for the nonmagnetic substrate grows monotonically. The observed at high strengths decrease of the magnetic substrate layer primary structure wave number may be due to the intensely forming in the strong field secondary and tertiary peaks, as if these were pushing the primary ones apart. Meanwhile, a nonmagnetic substrate has only the secondary peaks, and their formation with the stronger fields can only slow down the growth of the wave number upon field dependence.

Fig. 6 shows the experimental dependence of the instability supercritical wave number upon the fluid layer thickness. This dependence is monotonically decreasing.

The comparison shows that the magnetic and nonmagnetic substrates dependences are practically similar at great thicknesses. The difference exists only in the thin layers where the magnetic substrate layer wave number is greater than the nonmagnetic substrate one. This confirms the above conclusion that as the fluid layer thickens the effect of the substrate properties on the instability development process rapidly weakens.

## 3. Discussion

As noted above, the instability of the magnetic fluid layer on magnetizable substrate in perpendicular magnetic field had been theoretically discussed in [4–6]. These had obtained the different dispersion relation for the fluid layer free surface disturbances that let find and analyze the development of the most instable surface disturbances. The relevant calculation demonstrates the qualitative agreement of the obtained experimental regularities with the theoretical studies results. However, the quantitative results significantly differ. This is why here we will only discuss the most interesting criterion of the surface instability occurrence.

In [4–6] the conclusion is that the magnetic substrate must stimulate the instability development; such instability must occur at lower magnetic field strengths than with a nonmagnetic substrate; this completely agrees with the above conclusion based on the experimental data. Thus, in [6], the following expression was obtained for the magnetic fluid critical magnetization at which the nonmagnetic substrate layer instability begins developing:

$$M_c = 2\sqrt[4]{\rho g \sigma / \mu_0^2}, \qquad (1)$$

where $g$ is the free fall acceleration, $\mu_0$ the magnetic constant. The critical magnetization of a magnetic fluid on magnetizable substrate is determined by the following equality:

$$M_{cm} = \sqrt[4]{\rho g \sigma} \sqrt{\frac{2(\mu/\mu_0 + 1)}{\mu}}, \qquad (2)$$

where $\mu$ is the substrate material magnetic permeability supposed constant. From (1), (2) it is easily seen that $M_{cm}/M_c < 1$, i.e., the layer instability occurrence threshold with magnetic substrate is lower than for the nonmagnetic one. The relevant calculation shows that the critical field strength for the magnetic substrate layer is by about 30% lower than the nonmagnetic substrate one, other factors being equal. This relatively well agrees with the experimental (Fig. 2).

Thus, the studies have revealed a number of new regularities for the development of instability of the thin layer of magnetic fluid on magnetizable plate in perpendicular magnetic field. The magnetic substrate has been shown to bring down the instability threshold; with this, the effect of the substrate characteristics on the instability development regularities disappears as the magnetic fluid layer thickens. Certain regularities qualitatively agree with the forecasts of the existing theories; however, the quantitative results significantly differ. With this, the experimentally revealed effect of the secondary and tertiary peaks resulting from the instability development has, so far, no theoretical justification. These discrepancies may result from the nonlinear effects and require a deeper analysis. This means that further, both theoretical and experimental, studies are needed in this domain.

**Acknowledgments**
The authors thank prof. Yu.I. Dikansky for valuable discussions of the work results. This work was supported by the grant of the President of Russian Federation No. MK-6053.2012.2 and also by Ministry of Education and Science of the Russian Federation within the scientific program "Development of Scientific Potential of Higher School".

**Figures**

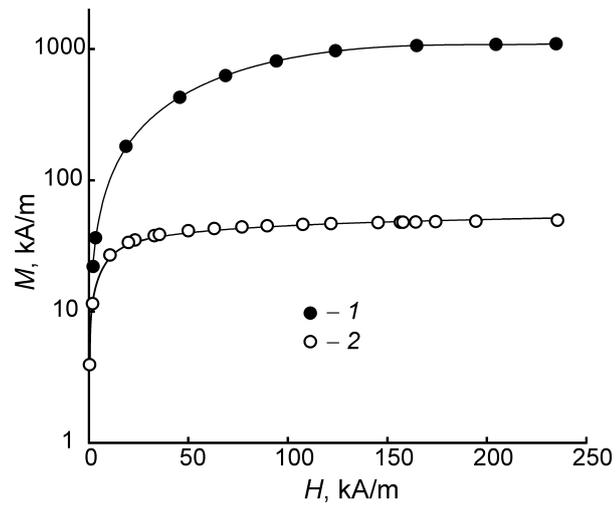

Fig. 1. Experimental magnetization curves for plate (*1*) and magnetic fluid (*2*) materials.

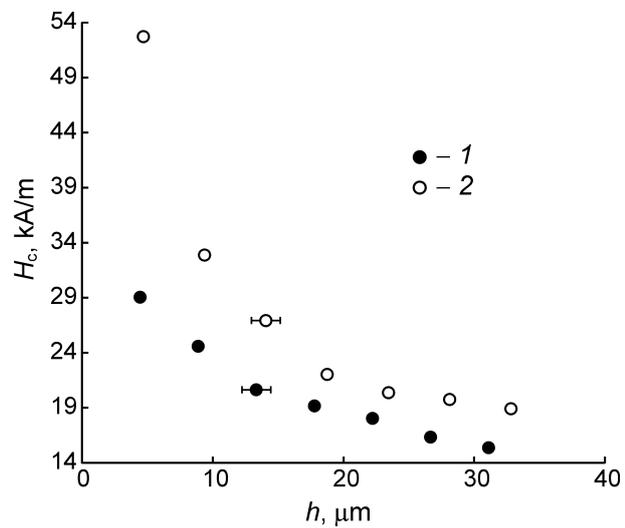

Fig. 2. Dependences of magnetic field critical strength upon ferrofluid layer thickness for magnetizable (*1*) and nonmagnetic (*2*) substrates.

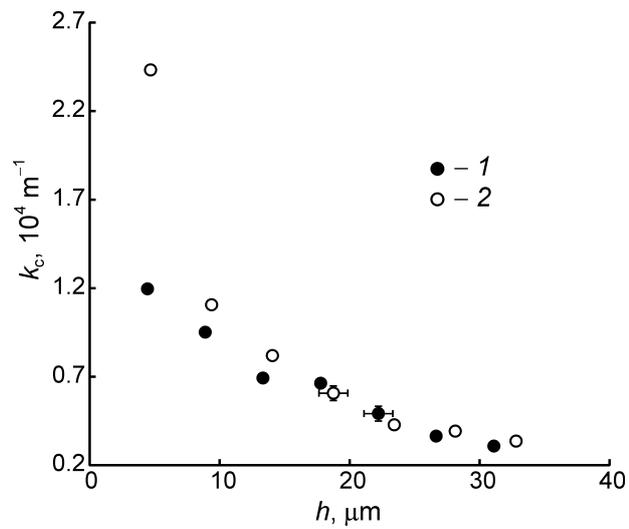

Fig. 3. Dependences of instability critical wave number upon fluid layer thickness for magnetizable (*1*) and nonmagnetic (*2*) substrates.

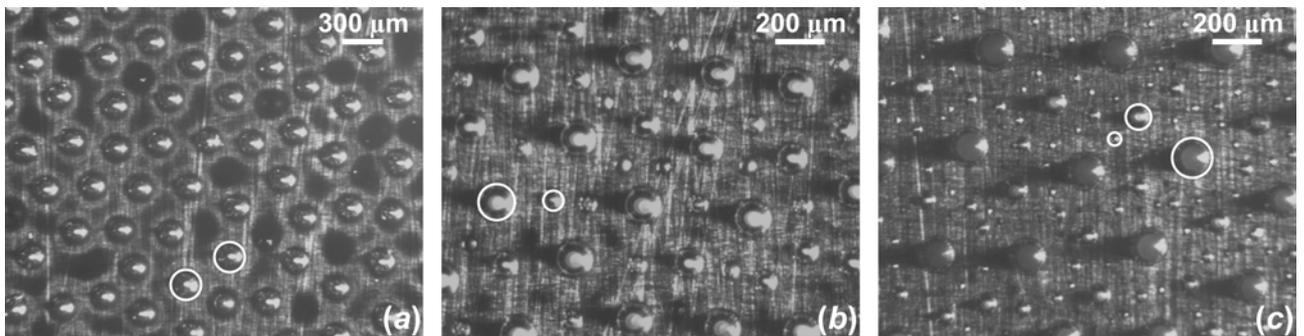

Fig. 4. Structure occurring at disintegration of ferrofluid layer on magnetic substrate at different external magnetic field strengths: (*a*) $H = 42$ kA/m; (*b*) $H = 93$ kA/m; (*c*) $H = 185$ kA/m. Layer thickness $h = 8$ μm.

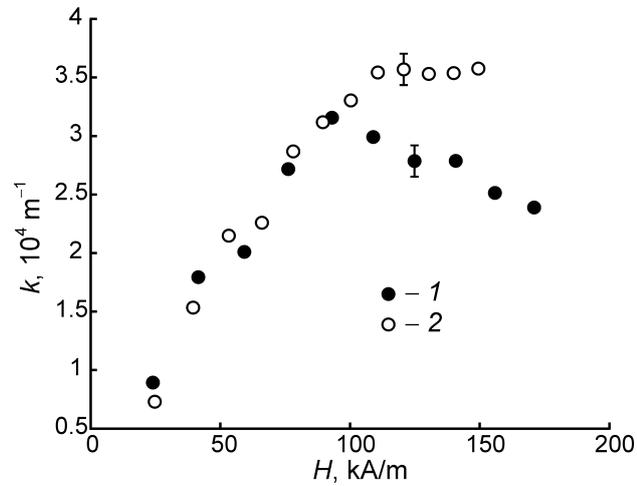

Fig. 5. Dependences of instability supercritical wave number upon magnetic field strength for magnetizable (*1*) and nonmagnetic (*2*) substrates. Layer thickness $h = 8$ μm.

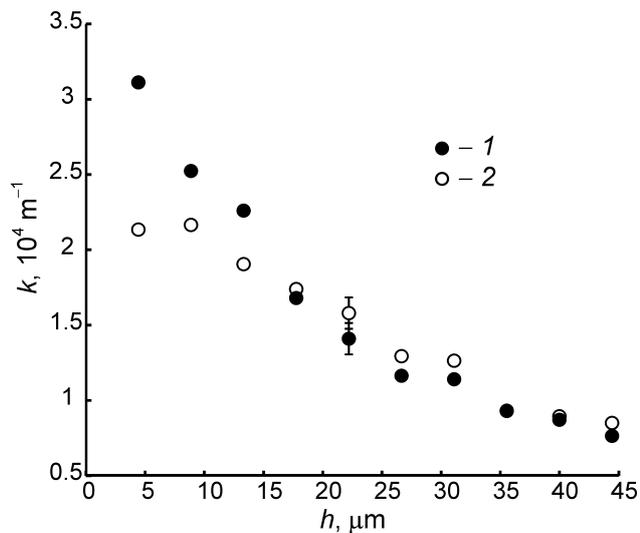

Fig. 6. Dependences of instability supercritical wave number upon fluid layer thickness for magnetizable (*1*) and nonmagnetic (*2*) substrates. External magnetic field strength $H = 50$ kA/m.